\documentclass[10pt,conference]{IEEEtran}

\newcommand\eg[1]{{\textit{e.g.}#1}}
\newcommand\etal[1]{{\textit{et al.}#1}}
\newcommand\ie[1]{{\textit{i.e.}#1}}
\usepackage{flushend}
\usepackage{hyperref}
\usepackage{graphicx}
\hypersetup{
    colorlinks=true,
    linkcolor=black,
    urlcolor=black,
}
\usepackage[shortcuts,acronym]{glossaries}
\newacronym{v2v}{V2V}{Vehicle-to-Vehicle}
\newacronym{v2x}{V2X}{Vehicle-to-Everything}
\newacronym{ntn}{NTN}{Non-Terrestrial Networks}
\newacronym{dsrc}{DSRC}{dedicated short-range communications}
\newacronym{uav}{UAV}{unmanned aerial vehicle}

\begin{document}

\title{\fontsize{16pt}{0pt} \textbf{Wireless Communication as an Information Sensor for Multi-agent \\ Cooperative Perception: A Survey}}

\author{Zhiying Song, Tenghui Xie, Fuxi Wen,~\textit{Senior Member,~IEEE}, Jun Li
\thanks{
The authors are with the School of Vehicle and Mobility, Tsinghua University. Email: \tt\small{wenfuxi@tsinghua.edu.cn}.}
}

\maketitle

\begin{abstract}
Cooperative perception extends the perception capabilities of autonomous vehicles by enabling multi-agent information sharing via Vehicle-to-Everything (V2X) communication. Unlike traditional onboard sensors, V2X acts as a dynamic ``information sensor'' characterized by limited communication, heterogeneity, mobility, and scalability. This survey provides a comprehensive review of recent advancements from the perspective of information-centric cooperative perception, focusing on three key dimensions: information representation, information fusion, and large-scale deployment. We categorize information representation into data-level, feature-level, and object-level schemes, and highlight emerging methods for reducing data volume and compressing messages under communication constraints. In information fusion, we explore techniques under both ideal and non-ideal conditions, including those addressing heterogeneity, localization errors, latency, and packet loss. Finally, we summarize system-level approaches to support scalability in dense traffic scenarios. Compared with existing surveys, this paper introduces a new perspective by treating V2X communication as an information sensor and emphasizing the challenges of deploying cooperative perception in real-world intelligent transportation systems.
\end{abstract}


\section{Introduction}
\label{sec:intro}
Autonomous vehicles rely on perception systems to navigate complex scenarios. These systems typically integrate multiple onboard sensors, such as LiDAR and cameras. However, single-vehicle perception has inherent limitations, such as sensor range constraints and occlusion, which fragment the operational design domain and compromise safety \cite{Wang2020}. 

To bridge these gaps, V2X communication offers a solution by functioning as a virtual \textit{information sensor}. V2X enables the ego vehicle to receive information from other agents via wireless communication. Unlike onboard sensors that have fixed scanning perspectives, V2X provides data from other agents' viewpoints, often complementing the ego vehicle's perception and filling in gaps caused by sensor limitations \cite{liu2023towards}.
This concept is known as {cooperative perception}, which leverages V2X communication to allow vehicles to share perceptual data, thereby extending their sensing range and enhancing system robustness \cite{caillot2022survey}.

\begin{figure}[t]
\centerline{\includegraphics[width=0.45\textwidth]{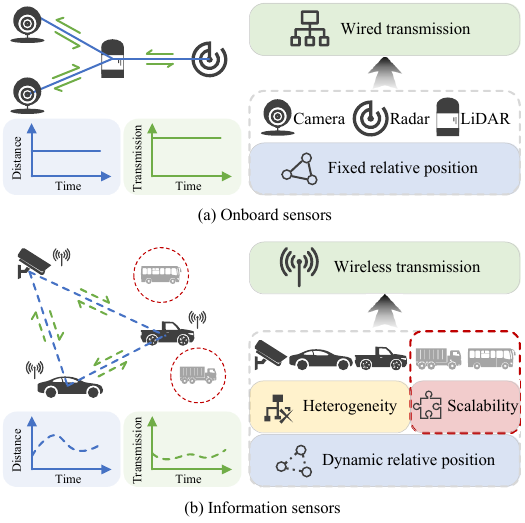}}
\vspace{-3mm}
\caption{Comparison of key characteristics between onboard sensors and information sensors.}
\vspace{-5mm}
\label{fig:info_sensor}
\end{figure}

The primary difference between onboard sensors and information sensors lies in their connection to the processing system (Figure \ref{fig:info_sensor}). Onboard sensors are physically integrated into the vehicle's electronic system using wired communication protocols. In contrast, information sensors rely on wireless communication to transmit data, which introduces several key characteristics. First, there is \textit{mobility}. Unlike fixed onboard sensors, both sending and receiving agents are mobile, and cooperative targets may change positions dynamically. This mobility imposes high precision requirements on the positioning systems of cooperative agents.
Second, there is \textit{heterogeneity}. While onboard sensors have fixed operating modes, cooperative agents vary widely in data content, model structures, and communication protocols. Third, there is a \textit{high dependence on communication}. Unlike the wired data transmission between onboard sensors, information must be transmitted in real-time over wireless networks, and its volume and quality are constrained by factors such as bandwidth and signal stability. 
Finally, there is \textit{scalability}. Unlike a fixed number of onboard sensors, information sensing involves a large number of agents whose network topology evolves in real-time. The number of participating agents can range from a few to hundreds, with information flowing between them.

The introduction of information sensors into cooperative perception systems presents several challenges. First, \textit{how to represent information}: encoding an agent's observation of the world into a form of information that meets communication constraints (\eg, real-time requirements and bandwidth limitations), overcomes agent heterogeneity, and supports large-scale deployment is a fundamental issue. Second, \textit{how to fuse information}: in complex and dynamic environments, the system must effectively integrate information while mitigating real-world noise, such as communication packet loss, latency, and localization errors. Third, \textit{how to schedule information}: in large-scale systems, planning the flow of information at the system level is crucial to ensure stable operation.

This survey focuses on these challenges, centering on information as the key element of cooperative perception systems. We review recent progress in three critical areas: information representation, information fusion, and large-scale deployment. We identify the key challenges, summarize the latest academic advancements, and propose potential research directions. Compared to existing surveys \cite{han2023collaborative,huang2023v2x,yazgan2024survey}, which primarily focus on information fusion, this paper introduces a novel perspective by emphasizing the role of information sensors and highlighting the importance of information representation and scalability in real-world implementations. We pay particular attention to the challenges that arise during the deployment of cooperative perception systems in intelligent transportation environments.

\section{Information Representation}
Information representation refers to the process of encoding information. High-quality information representation can optimize the utilization of communication and computational resources. In contrast, low-quality representations can overload the communication network, hindering the real-time performance of the system and potentially even causing the cooperative perception system to fail.

\subsection{Abstracting the World}
Cooperative perception involves two primary types of information: Cooperative Awareness Messages (CAM) and Cooperative Perception Messages (CPM). CAM conveys the vehicle's state, such as position and size, while CPM refers to the perception data generated by the vehicle's onboard sensors.

The simplest form of cooperation would involve vehicles sharing only CAM. In a fully connected IoT ecosystem, all traffic participants could exchange these messages, enabling a complete and precise awareness of the environment. That said, realizing this level of cooperation depends on extensive network penetration, a challenge that remains difficult to overcome in the foreseeable future.
Depending on the type of CPM, cooperative perception can be categorized into three approaches: data-, feature-, and object-level.
At the \textit{data level}, raw data such as camera images is directly exchanged between agents \cite{kim2014multivehicle,chen2019cooper}. Although this method preserves the most detailed information, it is not practical due to the substantial bandwidth requirements, which can overwhelm networks and cause packet loss and delays \cite{arnold2020cooperative}.
The \textit{feature-level} representation involves encoding raw data into more compact, low-dimensional representations using neural networks \cite{xiao2018multimedia}. This reduces the amount of data exchanged, while retaining key features necessary for perception \cite{Xu2022a}. Nevertheless, the black-box nature of neural networks makes standardization difficult, and the heterogeneity within the cooperative network complicates this further \cite{yazgan2024survey}.
At the \textit{object level}, high-level information such as the positions and poses of detected objects is shared \cite{Song2023}. This approach demands minimal bandwidth, reduces latency, and is easier to standardize and interpret, making it ideal for practical applications \cite{rauch2012car2x}. However, the abstraction of raw data into object-level information leads to significant information loss \cite{aeberhard2017object}. This limits the potential of cooperative data to be used for tasks beyond object detection, such as occupancy grid prediction \cite{song2024collaborative}, or for seamless integration into end-to-end driving frameworks \cite{cui2022coopernaut,yu2025end}.

In practical traffic systems, the choice of representation remains an open question, as none of the existing approaches are optimal across all dimensions, including communication requirements, information loss, and standardization. Finding a universally effective representation is the key challenge. We suggest two possible directions for future development. One approach is to develop explicit representations, abstracting the world into a form that minimizes information loss while remaining more compact and less heterogeneous than raw data. For example, drawing on recent advances in the 3D Gaussian domain \cite{kerbl20233d}, one can utilize 3D ellipsoids to encode environmental information \cite{li2024gmmap}. Another avenue involves exploring a universal feature space. The success of large vision-language models demonstrates that multi-modal data, such as text and images, can be aligned into a unified feature space \cite{radford2021learning}. This raises the possibility of a unified autonomous driving model that could extract multi-modal data into a shared feature space, addressing the heterogeneity and standardization issues of feature-level representations. 

\subsection{Reducing Information Volume}
Once the method of information representation is determined, the next step is to carefully consider what information to transmit. This is crucial because, in cooperative systems, communication bandwidth is extremely limited, and the effective transmission performance significantly decreases due to factors such as high vehicle mobility, complex traffic scenarios, and multi-source interference. 
Cui \etal~ found that the throughput of C-V2X, which is the mainstream standard used in cooperation, in real-world vehicle-to-vehicle cooperative scenarios is limited to less than 10 Mbps \cite{cui2022coopernaut}. This translates to a bandwidth that can only support the transmission of about 4.16 million pixels, 10 LiDAR points, or 4,800 64-channel depth features per second, which is far less than the data or intermediate features generated by onboard perception systems.
On the other hand, cooperative perception information often contains considerable spatial redundancy \cite{huwhere2comm}. By intelligently sparsifying and prioritizing critical perception data, it is possible to better adapt to limited communication resources and significantly improve the efficiency of cooperation \cite{huang2023v2x}.

Currently, one of the primary research directions focuses on optimizing redundant information, with a particular emphasis on object detection tasks.
In object-level cooperative perception, which primarily transmits structured information, a representative method is AICP \cite{zhou2022aicp}. It determines the importance of cooperative information based on factors such as distance, relative velocity, direction, and the category of objects at the object level, as well as message creation time and decay rate at the message level, and the overall informativeness of received messages at the vehicle level. 
For data-level representation, segmentation or clustering can be employed to extract sparse representations from dense raw data \cite{zhang2023robust}.
For example, Ding \etal~ propose a compact message unit called point cluster \cite{dingpoint}, using strategic sampling to preserve structural information with fewer key points for dense LiDAR.
For feature-level representation, a standard method is to integrate the information selection and object detection processes into an end-to-end learnable framework. For instance, Where2Comm \cite{huwhere2comm} trains the model to learn feature map masks based on classification confidence from an object detection model, thereby achieving better cooperative object detection accuracy. TransIFF \cite{chen2023transiff} directly encodes sensor data into semantic token sequences, selecting a subset of tokens to be transmitted, where the object detection task directly supervises the token encoding process.

Another direction focuses on driving tasks. Vehicles mainly depend on perception data from critical areas, such as the forward collision risk zone \cite{li2023transcendental}, while information from distant regions beyond lane boundaries is less critical. AutoCast \cite{qiu2022autocast} predicts future collision risks and selectively transmits point cloud data of the relevant objects. More recently, Plan2Comm \cite{xie2024towards} determines which features to share by predicting occupancy flow, generating costmaps, and using a redundancy-complementarity equalizer based on planned trajectories.

Despite notable progress, existing methods for information reduction still face several limitations. Most current approaches are tightly coupled with specific tasks such as object detection or trajectory planning, which may hinder their adaptability to diverse perception requirements. Moreover, these methods often rely on joint training using data from multiple agents to enable task-driven information selection, an assumption that is difficult to realize in real-world deployments.
Future research should aim to develop more general and task-adaptive information selection strategies that can operate effectively under varying communication budgets. In addition, more attention should be paid to quantifying the information needs of cooperative perception based solely on ego-vehicle observations, enabling the vehicle to request only the most relevant external information actively.

\subsection{Compressing and Conveying the Messages}
Compression and encoding of cooperative information are core modules for cooperative perception systems, as their performance directly determines whether perception information can be transmitted within a limited communication bandwidth while preserving semantic integrity. 

For data-level representation, point cloud or image compression algorithms can be applied to compress relevant information by entropy coding low-level data such as geometric coordinates and color attributes \cite{van2019image}. Nevertheless, applying these methods to deep feature representations often fails to achieve high compression ratios.
At the feature level, the mainstream approach is to use neural networks for compression. For example, V2VNet~\cite{Wang2020} uses convolutional neural networks. In recent years, deep generative models such as autoencoders \cite{hinton2006reducing} and their variations \cite{kingma2013auto,he2022masked} have made significant progress. These models encode raw perception data into a low-dimensional latent space, generating compact feature tensors for transmission and reconstructing semantic information through decoder networks. 
In cooperative perception, OPV2V~\cite{Xu2022a} was among the first to introduce autoencoders for feature compression and decompression. 
By leveraging high-order statistical properties and semantic correlations, these generative models achieve higher compression efficiency while maintaining comparable reconstruction quality. To further reduce the communication overhead of already compressed low-dimensional features, V2VNet adopts traditional floating-point quantization and entropy encoding, whereas Hu~\etal~~\cite{hu2024communication} introduce a task-driven codebook that maps feature maps to integer indices by minimizing reconstruction error.

Overall, existing neural methods inherently adopt a lossy compression paradigm. At high compression ratios, they often cause severe semantic distortion: once compression exceeds a critical threshold, the loss of key perception features, such as small object contours or motion parameters, leads to a sharp decline in downstream task performance. This issue is systematically demonstrated by Liu~\etal~~\cite{liu2023towards}, who evaluated typical methods across three datasets under different compression ratios. The results show that as the compression ratio increases (\ie, more information is lost), cooperative perception accuracy degrades rapidly, even falling below that of single-vehicle perception. Developing compression and codec methods that are efficient, information-preserving, and easy to standardize will be a key future research direction.

\section{Information Fusion}
Information fusion refers to the process in which a collaborative vehicle decodes and reconstructs messages received from other agents, and then integrates them with its own information. Existing studies typically investigate the fusion process under two types of conditions:
a) \textit{Ideal conditions}: These assume homogeneous data sources as well as perfect communication, localization, and perception capabilities among collaborating vehicles. Research under this setting aims to explore the theoretical upper bounds of cooperative system performance.
b) \textit{Non-ideal conditions}: These involve imperfections such as heterogeneity, communication delays, localization errors, or packet loss. In such cases, research focuses on ensuring the performance lower bound of cooperative systems in real-world environments.

\subsection{Ideal Condition}
A straightforward approach is to directly aggregate the aligned perception results from multiple vehicles using operations such as summation, maximum, or average. For instance, Cooper~\cite{Chen2019b} concatenates the point clouds from two vehicles and performs object detection on the combined data. F-Cooper~\cite{Chen2019} extends this to the feature level by first extracting deep features from point clouds on each vehicle and then applying max pooling across features from multiple agents. These methods are simple and symmetric, avoiding bias among agents during training. However, when there is asymmetry in the data among agents, such as from heterogeneous sensors, these simple methods become ineffective. They cannot account for variations in data distribution, which may lead to suboptimal fusion and degraded perception performance.

To overcome these limitations, several advanced information fusion methods have been proposed. 
V2VNet~\cite{Wang2020} uses a GNN-based cross-agent aggregation module. Each agent updates its features by dynamically aggregating messages from neighbors via spatial transformation and temporal warping.
OPV2V~\cite{Xu2022a} applies self-attention to reason about inter-agent interactions at the same spatial locations.
V2X-ViT~\cite{Xu2022} models inter-agent importance by computing attention based on agent and edge types, and introduces a multiscale window attention module to aggregate features from multiple agents using windows of varying sizes, thereby enhancing robustness to spatial misalignments. Under ideal conditions, existing research has achieved high performance; however, the greater challenge lies in addressing information fusion under real-world conditions, which are often characterized by various imperfections \cite{yazgan2024survey}.

\subsection{Heterogeneity}
As introduced in Section~\ref{sec:intro}, heterogeneity is a fundamental characteristic of information sensors, manifesting in both model structures and sensor data formats. Model-level heterogeneity refers to differences in intermediate representations generated by different models, which must be aligned before meaningful fusion can occur. Data-level heterogeneity arises from combining diverse sensor modalities or from variations in sensor configurations within the same modality (\eg, LiDAR with different point densities).

To address model heterogeneity, MPDA~\cite{xu2023bridging} introduces a learnable feature resizer that aligns multi-dimensional representations by adjusting spatial resolutions and channel depths. It further employs a sparse cross-domain transformer to extract domain-invariant features, enabling collaboration between agents without exposing internal model details. PnPDA~\cite{luo2024plug} tackles the same problem using a semantic calibration framework based on contrastive learning. It consists of a semantic converter and enhancer to map model-specific features to a shared semantic space, allowing new models to join collaborative tasks without altering existing architectures.
For data heterogeneity, the challenge of fusing different sensor modalities is similar to that in single-vehicle perception. In the context of cooperative perception, HM-ViT~\cite{xiang2023hm} addresses this by employing modality-specific encoders to generate BEV features, which are then fused via a heterogeneous 3D graph transformer with local and global attention, enabling effective integration of camera and LiDAR data in heterogeneous V2V scenarios. For intra-modality heterogeneity, \ie, differences in sensor configurations within the same modality, research remains limited in the V2X domain. But datasets like HeLiPR~\cite{jung2024helipr} have recently emerged in single-vehicle settings, drawing growing attention to this issue.
HEAL~\cite{lu2024extensible} tackles both model and data heterogeneity. It first establishes a unified feature space through a Pyramid Fusion network for existing agents. When new agents join, a backward alignment mechanism adapts only their front-end encoders while keeping the shared fusion and detection modules fixed, thus maintaining compatibility with minimal retraining cost.

While existing studies have made initial progress in addressing heterogeneity, they still fall short of being applicable in real-world scenarios. A significant limitation is that most current approaches assume all participating agents can share data and models for joint training, which is impractical in realistic deployments due to constraints on privacy, bandwidth, and system autonomy. In practice, we envision a more flexible paradigm: vehicles should be able to collaborate on the fly with any encountered agents, regardless of underlying data or model heterogeneity, without requiring prior joint training. How to pave the way toward such a flexible and scalable cooperative framework remains an open and challenging question.

\subsection{Imperfect Communications}
Unreliable communication in cooperative perception primarily manifests as transmission latency and packet loss.

In cooperative perception systems, latency arises from data collection, perception inference and wireless transmission. Such delays can mislead fusion algorithms, causing perceptual inconsistencies and even degrading performance below that of single-vehicle perception~\cite{aoki2022time}. Hence, latency compensation is a critical component.
A common strategy involves predicting features from past frames and fusing them with the current frame. For example, SyncNet~\cite{lei2022latency} utilizes LSTM-based prediction and attention-based fusion to handle delayed messages, whereas FFNet \cite{yu2023vehicle} leverages temporal continuity to forecast future deep features. 
Alternative approaches include CNN-based modeling in V2VNet \cite{Wang2020} and attention-based alignment in V2X-ViT \cite{Xu2022}. However, these methods exhibit degraded performance under longer latency~\cite{yu2023vehicle}. More recently, TraF-Align~\cite{song2025traf} proposes an end-to-end framework that learns feature-level trajectories from past observations. By generating temporally ordered sampling points along these trajectories, the model directs attention from current-time queries to relevant historical features. This enables temporal alignment, corrects spatial misalignments, and promotes semantic consistency across agents. Nevertheless, TraF-Align depends heavily on accurate trajectory prediction and attention learning. 

In real-world scenarios, V2X communication often suffers from packet loss due to signal power fluctuations caused by factors such as rapid vehicle movement, harsh weather, and potential malicious interference \cite{huang2023v2x}. As seen in AutoCast \cite{qiu2022autocast}, a straightforward strategy is extrapolating missing data packets. To more effectively address this challenge, Li~\etal~\cite{li2023learning} proposed a communication-aware recovery network that reconstructs corrupted shared features using a tensor-wise filtering scheme. They further introduced a V2V attention module that integrates intra-vehicle attention with uncertainty-aware inter-vehicle attention, thereby improving the robustness of feature fusion under unreliable transmission. Complementarily, Ren~\etal~\cite{ren2024interruption} designed a communication-adaptive spatiotemporal prediction model that infers missing information based on historical cooperative data. To enhance the model’s adaptability, they incorporated knowledge distillation and a curriculum learning framework, improving overall system performance in the presence of packet loss.

In summary, while several studies have explored solutions for imperfect communication, such as latency and packet loss, these approaches often rely on neural networks, which may face generalization challenges in real-world deployments. Further research is needed to improve their robustness and applicability in practical scenarios.

\subsection{Pose Errors}
Pose information is used to estimate the coordinate transformation matrix between cooperative vehicles, thereby unifying their local reference frames. In practice, this matrix is typically derived from localization systems. However, measurement errors and sensor noise lead to spatial misalignment between agents, which severely degrades the effectiveness of shared information. Current research explores three main strategies to address this issue: a) Improving localization accuracy through better sensors and algorithms, b) Aligning agents to a shared reference map, often a high-definition map, via map-matching techniques, c) Performing online spatial calibration directly from collaborative data. This approach is particularly relevant to cooperative perception and can be further categorized according to information representations.

Early studies utilized occupancy maps constructed from radar or cameras and aligned them via map registration~\cite{Kim2014}, but suffered from large transformation errors due to sensor limitations. With the advancement of LiDAR technology, point cloud registration has become a dominant method for data processing. Coarse registration techniques, such as Fast Point Feature Histograms~\cite {rusu2009fast}, are used for initial alignment, followed by fine registration using ICP and its variants~\cite{Besl1992}. Although effective in SLAM and local sensor calibration, these methods are often impractical for V2X applications due to bandwidth constraints in sharing raw sensor data.
To reduce communication overhead, feature-level alignment methods have emerged. Vadivelu \etal~~\cite{vadivelu2021learning} propose a deep neural network to regress relative poses from intermediate perception features, though it requires ground-truth poses for training. CoAlign~\cite{lu2023robust} bypasses this need by learning consistent pose graphs across agents. 
TrajMatch~\cite{ren2023trajmatch} utilizes tracked object trajectories and semantic features from detectors to align multiple roadside LiDARs. However, these methods often require consistent descriptors across agents and are sensitive to hyperparameters and environmental changes.
Object-level alignment techniques transfer only structured information and solve spatial alignment by establishing correspondences among perceived objects. OptiMatch~\cite{Song2023} formulates the alignment as an optimal transport problem between sets of detected object centers, minimizing transport cost via the Sinkhorn algorithm. VIPS~\cite{shi2022vips} builds a graph from geometric relationships among detected objects and solves the alignment through graph matching. Both methods require absolute coordinates and are sensitive to detection noise. CBM~\cite{song2023spatial} addresses these issues by encoding only intra-set relative geometry and introducing error compensation during alignment.

Current studies investigating the impact of pose errors typically simulate them by injecting predefined noise (commonly Gaussian) into well-calibrated datasets. However, such datasets are often carefully processed and offer high pose precision, which limits the realism of the evaluation. Future research should focus on collecting real-world localization data under cooperative scenarios and developing models that are designed and tested on these more realistic datasets. This would enable the advancement of cooperative perception systems that are more robust to real-world pose inaccuracies.

\section{Large-scale Deployment}
Scalability is a fundamental requirement for the large-scale deployment of cooperative perception systems. It defines the system’s ability to maintain stable performance as the number of participating vehicles increases. Specifically, it affects communication efficiency, which describes how bandwidth is distributed as the number of agents grows; computational load, which refers to whether processing costs remain manageable in a distributed setting; and real-time capability, which concerns whether low-latency operation can still be maintained as system complexity increases.

As connectivity expands in large-scale V2X networks, designing cooperative mechanisms that ensure scalability becomes increasingly important. This raises several critical questions: Who should each agent communicate with? When should communication occur? What information should be transmitted? And who should make these decisions?

From a system architecture perspective, several approaches have been proposed to support scalability. EMP~\cite{zhang2021emp} adopts an edge-assisted design in which vehicles upload raw sensor data to edge servers for centralized fusion and inference. To reduce communication overhead, EMP employs Voronoi partitioning to restrict uploads to relevant spatial areas, and power diagrams to adapt these partitions based on available bandwidth. Data upload is scheduled using Delaunay triangulation, which terminates transmission once sufficient information is received. EMP supports real-time operation for up to six vehicles, with an average end-to-end delay of 93 milliseconds, and outperforms traditional V2V sharing schemes.
AutoCast~\cite{qiu2022autocast} presents a fully decentralized architecture that does not rely on infrastructure. It separates the system into control and data modules. The control module is responsible for exchanging metadata such as trajectories and object maps to define communication domains, while the data module handles LiDAR processing and planning. The transmission policy is modeled as a Markov decision process, with a greedy scheduler determining message priority based on object visibility and relevance. AutoCast demonstrates scalability in dense scenarios involving up to 40 vehicles.
Harbor~\cite{zhu2024boosting} proposes a hybrid architecture that integrates both V2V and V2I communication. Vehicles are designated as helpers or requesters, where helpers upload data on behalf of requesters to the edge server. The helper-requester assignment is modeled as a bandwidth-aware bipartite graph optimization problem, considering factors such as mobility and network connectivity. A polynomial-time approximation algorithm is used to ensure efficient pairings. Deadline-awareness at the application layer and priority scheduling at the MAC layer help guarantee real-time performance. Harbor achieves latency reductions of 18\% to 57\% and detection accuracy improvements of up to 36\% compared to pure V2V or V2I approaches.

In addition to architectural design, intelligent information scheduling is another key to scalability. For example, Select2Coll~\cite{liu2024select2col} models the spatiotemporal importance of semantic information using a graph neural network to identify contributive collaborators and reduce redundant communication. Chiu~\etal~\cite{chiu2023selective} propose a two-stage communication strategy where only 2D object center coordinates are exchanged in the first round, followed by selective sharing of point-level features based on utility scores in the second. Who2Com~\cite{liu2020who2com} introduces a three-stage handshake mechanism in which agents first send compressed requests, receive relevance scores from others, and then select collaborators accordingly. When2Com~\cite{liu2020when2com} focuses on learning both communication grouping and timing, using self-attention mechanisms to prune weak connections and determine optimal communication moments, thereby improving efficiency.

Despite these advancements, several limitations remain. Many designs assume homogeneous sensor configurations, which may not reflect real-world heterogeneity. Furthermore, large-scale experiments are often conducted in simulation, as collecting real-world multi-agent cooperative perception data is expensive. Moreover, the absence of standardized benchmarks for evaluating scalability significantly limits the comparability and progress of current methods. Future research should explore more generalizable architectural designs that are robust to sensor and agent diversity. It is also essential to develop realistic large-scale datasets and unified evaluation protocols to better measure and accelerate the development of scalable cooperative perception systems.

\section{Conclusion}
This survey reviews cooperative perception from a sensor information perspective, highlighting V2X communication challenges. We summarize progress in representation, fusion, and deployment, and note key limitations such as task-specific designs, joint training dependence, and poor scalability. Future work should focus on generalizable, robust, and scalable frameworks, along with realistic datasets and standardized benchmarks to bridge the gap to practical deployment.

\bibliographystyle{IEEEtran}
\bibliography{reference}

\end{document}